\documentclass[pra,aps,epsfig,twocolumn,amsmath,amssymb]{revtex4}

\newcommand\be{\begin{eqnarray}}
\newcommand\ee{\end{eqnarray}}
\newcommand\nn{\nonumber}

\usepackage{graphicx}

\begin{document}
\bibliographystyle{prsty}
\title{Quantum theory of resonantly enhanced four-wave mixing: mean-field and
exact numerical solutions}
\author{Mattias T. Johnsson and Michael Fleischhauer}
\affiliation{Fachbereich Physik, Universit\"{a}t Kaiserslautern, D-67663 Kaiserslautern,
Germany}
\date{\today}

\begin{abstract}
We present a full quantum analysis of resonant forward 
four-wave mixing based on
electromagnetically induced transparency (EIT).
In particular, we study the regime of efficient nonlinear conversion
with low-intensity fields that has been predicted from a semiclassical analysis.
We derive an effective nonlinear interaction Hamiltonian in the adiabatic
limit. In contrast to conventional nonlinear optics this Hamiltonian
does not have a power expansion in the fields and the conversion
length increases with the input power.
We analyze the stationary wave-mixing process in the forward scattering
configuration using an exact numerical analysis for up
to $10^3$ input photons and compare the results with a mean-field approach.
Due to quantum effects, complete conversion from the two pump fields
into the signal and idler modes is achieved only asymptotically
for large coherent pump intensities or for  pump fields in 
few-photon Fock states.
The signal and idler fields are perfectly quantum correlated 
which has potential 
applications in quantum communication schemes. 
We also discuss the implementation of
a single-photon phase gate for continuous quantum computation.
\end{abstract}


\maketitle


\section{Introduction}

The cancellation of linear absorption and refraction in resonant
atomic systems by means of electromagnetically induced transparency
(EIT) \cite{harris1997,lukin2001,marangos1998} 
has led over the past 10 years to fascinating 
developments
in nonlinear optics \cite{Harris1990,Stoicheff1990}.
For example, coherently driven, resonant atomic vapors
under conditions of EIT allow for
complete frequency conversion in distances so short that phase
matching requirements become irrelevant
\cite{jain1996}. It has been predicted that resonantly enhanced
nonlinear interactions of light and atoms based on EIT
will lead to a new regime of efficient nonlinear
optics on the level of few photons 
\cite{harris1998,harris1999,imamoglu2000}. 
Besides being of interest
in its own right, such a regime would be very important
for applications in quantum communication and information processing.
It is clear that quantum effects will play an essential rule in this regime
and the quantum dynamics may substantially deviate from semiclassical
predictions. With the exception
of a few exactly solvable problems, for example the resonantly enhanced
Kerr effect \cite{schmidt1996,imamoglu2000}, 
quantum treatments of EIT-based nonlinear optics
have so far been restricted to small-fluctuation approximations.
In this paper we present a full quantum analysis of a particular
EIT-based nonlinear system, namely resonantly enhanced
four-wave mixing in a double-$\Lambda$ system
with co-propagating pump modes
\cite{hemmer1994,babin1996,lukin1997,popov1997,lu1998}.

Within a semiclassical analysis it has been shown that resonantly enhanced
four-wave mixing can lead to
complete conversion of the pump-field energy into the signal and idler modes
even for very weak pump fields. If the
atomic degrees of freedom can be eliminated adiabatically and losses can be
ignored, the semiclassical nonlinear problem is exactly integrable
\cite{korsunsky1999,korsunsky2002}.
For counter-propagating pump modes a phase transition to mirrorless
oscillations has been predicted 
\cite{lukin1998,review,fleischhauer2000book}
and experimentally verified \cite{zibrov1999}.
A linear fluctuation analysis has shown that
close to the threshold of oscillation an almost
perfect suppression of quantum noise
of one quadrature amplitude of a combination mode of the generated
fields occurs \cite{yuen1979,lukin1999}. In addition, 
sufficiently above threshold,
light fields with beat-frequencies tightly locked to the
atomic Raman-transition
and extremely low relative bandwidth
are generated \cite{fleischhauer2000}.

Assuming conditions of adiabatic following and considering the limit of
an infinitely long lived ground state coherence
we here derive a classical effective
nonlinear Hamiltonian which only contains field degrees of
freedom. In contrast to conventional four-wave mixing 
\cite{bloembergen1965,boyd1992,yuen1979}, this
Hamiltonian is a ratio of polynomial expressions and has no power expansion
in the fields. As a consequence
the conversion length increases rather than decreases with growing input
power. The evolution corresponding to this classical Hamiltonian can be
mapped to a single nonlinear pendulum and can be solved exactly.
However, the initial state of the pendulum
corresponding to vacuum in signal and idler modes is
an unstable equilibrium point.
Thus the initial evolution is entirely governed by quantum fluctuations.
Replacing the classical field variables in the effective Hamiltonian
by operators in normal ordered expressions, we obtain a
quantized Hamiltonian. Due to its nonpolynomial character it is not possible
to apply phase space techniques to study the quantum evolution of the 
fields.
Instead, making use of constants of motion, the stationary four-mode interaction
is reduced to a single-mode problem, which can be solved numerically
for up to $10^3$ input photons. 

We find the quantum dynamics to be
significantly different from the semiclassical prediction. In particular,
complete conversion is achieved only for input fields in a few-photon
Fock state or asymptotically for a very large coherent 
pump. The main features
of the dynamics, such as conversion efficiency and the dominant
oscillation frequency are reproduced by an appropriate 
mean-field theory which takes into account anomalous
correlations. Finally the quantum
statistics and correlations of the fields are analyzed and potential
applications for quantum communication and information processing discussed.
For example, we show that the resonant four-wave mixing process 
is an excellent source of quantum correlated photon pairs and 
can be used as a single-photon
phase gate for continuous quantum computation.


\section{Elimination of ac-Stark induced nonlinear refraction and
derivation of an effective Hamiltonian
\label{secAdiabaticElimination}}

\begin{figure}[ht]
\begin{center}
\includegraphics[width=5cm]{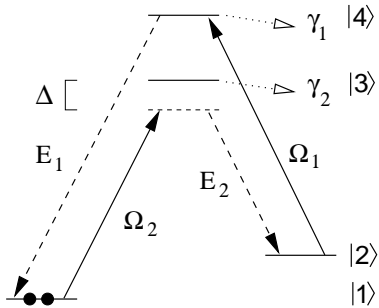}
\caption{Atoms in a double-$\Lambda$ configuration interacting
with two driving fields ($\Omega_{1,2}$) and two generated fields ($E_{1,2}$).}
\label{system}
\end{center}
\end{figure}

The standard resonant four-wave mixing scheme in a double-$\Lambda$
system is shown in Figure~\ref{system}.

The interacting beams consist of two pump fields with frequencies
$\nu_{d1}$ and $\nu_{d2}$ and Rabi-frequencies $\Omega_{1}$ and 
$\Omega_{2}$,
and two generated fields (signal and idler)
described by the
complex Rabi-frequencies $E_1$ and $E_2$,
with carrier frequencies
$\nu_1 = \nu_{\rm d1}+
\omega_0$ and $\nu_2=\nu_{\rm d2}-\omega_0$,
where $\omega_0=\omega_{\rm 2}-
\omega_{\rm 1}$ is the ground-state frequency splitting.
It is assumed here that the pump and generated fields are
pairwise in two-photon resonance and thus in four-photon resonance.
The latter is a consequence of energy conservation.
The two photon resonance is a consequence of phase matching and
semiclassical treatments show that signal and idler fields are
generated precisely at those frequencies. The
fields interact via the long-lived coherence on the
dipole-forbidden transition between the metastable ground states
$|1\rangle$ and $|2\rangle$.

\begin{figure}[ht]
\begin{center}
  \includegraphics[width=5cm]{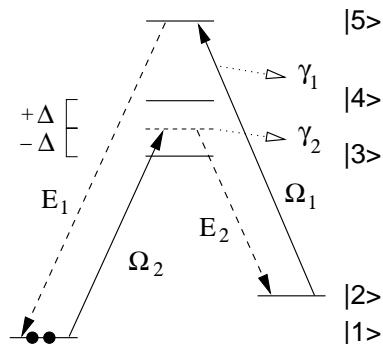}
  \caption{Modified double-$\Lambda$ system. Addition of a fifth level
allows the cancellation of destructive phase-shifts.}
\label{fig5level}
\end{center}
\end{figure}

The problem with this model is that associated with the finite detuning
$\Delta$, which is necessary to minimize linear absorption, are ac-Stark induced
nonlinear phase shifts. These phase shifts 
reduce conversion efficiency from the pump to the generated
modes, and at the same time increase the distance required for
conversion to take place \cite{johnsson2002}. As was shown in 
\cite{johnsson2002} these problems can be overcome by modifying
the system slightly. Instead of the original four-level scheme
a five-level set-up depicted in
Figure~\ref{fig5level} is used. This symmetric configuration cancels 
ac-Stark shifts.
In order to maintain the nonlinear interaction, which is also an odd function
of the detuning, it is however necessary
to choose atomic states such that the coupling
constant for one of the four transitions
$|1\rangle  \rightarrow  |3\rangle$, 
$|1\rangle  \rightarrow  |4\rangle$,
$|2\rangle  \rightarrow  |3\rangle$, and 
$|2\rangle  \rightarrow  |4\rangle$
has a different sign to the other three. This can easily
be accomplished by
using different hyperfine levels \cite{johnsson2002}.

Choosing the driving field $\Omega_1$ to be in resonance with the
$|2\rangle \to |5\rangle$ transition while the second driving field
$\Omega_2$ has a detuning $\pm\Delta$ with 
$|\Delta|\gg |\Omega_2|$ ensures that the linear losses
due to single-photon absorption are minimized.

The interaction of the fields with an ensemble of five-level atoms
is described by Maxwell's equations for the fields and a set of
density matrix
equations for the atoms which determine the atomic polarizations. 
Under conditions of adiabatic following and for the case of {\it classical} 
fields the latter are ususally solved
in steady-state and feed back into the first, yielding nonlinear field
equations of motion. This procedure is rather involved, particularly if
several atomic levels need to be taken into account, and is completely
inadequate for quantized electromagnetic fields. Consequently in this
paper we adopt a different procedure.We first derive an effective {\it
classical} interaction 
hamiltonian in the adiabatic limit and then quantize the effective theory
by replacing the field amplitudes by properly ordered operator expressions.

In order to rigorously calculate the medium response to classical fields,
one would have to solve the atomic density matrix equations
to all orders in all fields taking into
account all relaxation mechanisms. Instead we use a simplified
open-system model which allows to derive rather 
compact expressions for the atomic susceptibilities. In this model
the interaction of an individual atom with four classical modes is described
by a complex Hamiltonian, which, in a rotating-wave approximation corresponding
to slowly-varying amplitudes of the basis
$(|1\rangle \; |2\rangle \; |3\rangle \; |4\rangle \; |5\rangle)^T $,
can be written as
\begin{equation}
H_{\rm int} = -\hbar\left[
   \begin{matrix}
	0 & 0 & \Omega_2^* & \Omega_2^* & E_1^* \cr
        0 & 0 &  E_2^* & -E_2^* & \Omega_1^* \cr
        \Omega_2 & E_2 & -\Delta-i\gamma_2 & 0 & 0 \cr
        \Omega_2 & -E_2 & 0 & \Delta-i\gamma_2 & 0 \cr
 	E_1 & \Omega_1 & 0 & 0 & -i\gamma_1
   \end{matrix}\right].
\label{eqH5level}
\end{equation}
At the input the signal and idler modes $E_1$ and $E_2$ are assumed to have
zero amplitude and all atoms are in state $|1\rangle$. Taking into account
that optical pumping out of $|1\rangle$ is negligible as
$\Delta\gg |\Omega_2|$, this is a stable configuration corresponding to an
approximate adiabatic eigenstate (to lowest order in $\Delta^{-1}$) of $H$.
We now assume that the interaction takes place over sufficiently long time
scales, ensuring
that the atoms always stay in this approximate adiabatic eigenstate. Under
these conditions the Hamiltonian can be replaced by the corresponding
eigenvalue $\lambda_0$. Solving the 4th order characteristic equations
for the eigenvalues of (\ref{eqH5level}) and expanding the solutions in a power
series of $(\Omega/\Delta)$, with $\Omega$ being a characteristic value
of the Rabi-frequencies, one finds to lowest order
\begin{equation}
H_{\rm int}\longrightarrow \lambda_0 = \frac{\hbar}{\Delta}\left[\frac{
\Omega_1^*\Omega_2^*E_1 E_2+\Omega_1\Omega_2E_1^* E_2^*}
{|\Omega_1|^2+|E_1|^2} \right].
\label{eqlambda0}
\end{equation}
One recognizes two interesting and unusual features. First the eigenvalue
and correspondingly the medium polarization cannot be expressed as
a polynomial in the field amplitudes. Thus the resonant interaction
corresponds to an all-order nonlinear process. Secondly
despite the resonant interaction $\lambda_0$
has no imaginary component and there are hence no linear losses. This is
due to quantum interference associated with EIT. It allows for
efficient nonlinear interactions close to resonance without suffering from
linear absorption.

To quantize the interaction problem we replace the complex amplitudes
of the Rabi-frequencies in (\ref{eqlambda0}) by positive and negative
frequency components
of corresponding operators, choosing normal order in the
numerator and the denominator, multiply by the density of atoms
and integrate over the interaction volume. We thus arrive at the
effective interaction Hamiltonian between the quantized electromagnetic
fields
\begin{equation}
\hat H_{\rm int} = \frac{\hbar N A}{\Delta} \int\! {\rm d}z\, 
\left[\frac{\hat E_1^\dagger
\hat E_2^\dagger \hat\Omega_1\hat\Omega_2 +
\hat\Omega_1\hat\Omega_2\hat E_1\hat E_2}
{\hat\Omega_1^\dagger\hat\Omega_1 +\hat E_1^\dagger \hat E_1}\right]
\label{eqHint}
\end{equation}
where $N$ is the number density of atoms, $A$ the effective cross 
section of the beams, 
and $\hat E_1(z,t) =d_{51}
\sum_k \sqrt{{\nu_1}/{2\hbar\varepsilon_0 V}}\,  \hat a_{1k}(t)
{\rm e}^{-i (\nu_k-\nu_{1}) z/c}$ is the slowly varying positive frequency
operator of the signal Rabi-frequency. Correspondingly
$\hat \Omega_1(z,t) =d_{52}
\sum_k \sqrt{{\nu_{d1}}/{2\hbar\varepsilon_0 V}}\,  \hat b_{1k}(t)
{\rm e}^{-i (\nu_k-\nu_{d1}) z/c}$ denotes the slowly varying
positive frequency operator of the first pump Rabi-frequency.
The operators $\hat a_\mu$ and $\hat b_\mu$ obey harmonic oscillator
commutation relations. $V$ is the quantization volume, which
shall be identified with the interaction volume and $d_{ij}$ is the 
dipole
matrix element of the $|i\rangle \rightarrow |j\rangle$ transition.
It should be noted that numerator and
denominator commute and thus there is no ambiguity with respect to the
ordering of the two terms.

One can easily verify that there are four independent
quantities that commute with $\hat{H}_{\rm int}$ and are therefore constants of motion:
\begin{eqnarray}
\hat \Omega_1^\dagger \hat \Omega_1 + \hat E_1^\dagger\hat E_1 &=&
\, {\rm constant}, \label{eqCOM1} \\
\hat \Omega_2^\dagger \hat \Omega_2 + \hat E_2^\dagger\hat E_2 &=&
\, {\rm constant}, \label{eqCOM2} \\
\hat \Omega_1^\dagger \hat \Omega_1 -\hat \Omega_2^\dagger \hat \Omega_2
&=& \, {\rm constant}, \label{eqCOM3} \\
\hat \Omega_1^\dagger \hat \Omega_2^{\dagger} \hat E_1 \hat E_2  +
\hat \Omega_1 \hat \Omega_2 \hat E_1^{\dagger} \hat E_2^{\dagger}
&=&
\, {\rm constant}. \label{eqCOM4}
\end{eqnarray}
These are the quantum analogs of the classical Manley-Rowe relations
which express energy conservation in the system plus an equation expressing 
the conservation of the relative phase between the fields 
\cite{bloembergen1965}. 
The existence of these
constants of motion will considerably simplify the analysis.

It should be noted that the existence of these constants of motion are
not artifacts caused by using an effective rather than the full
Hamiltonian. We have also pursued a more rigorous derivation, which
involved writing the Heisenberg equations of motion for the atomic and
field subsystems seperately and including decay terms. In the limit
$\gamma \ll \Delta$ and negligible decay from the two ground states
this approach also yields (\ref{eqCOM1}) --
(\ref{eqCOM3}). Unfortunately the equations of  motion obtained in
this way have a number of unpleasant features and are not ameanable to
analytic solution, which is why we have chosen to use the effective
Hamiltonian (\ref{eqHint}) as the starting point of our discussion.


\section{Classical solutions for forward four-wave mixing}


To obtain classical solutions we use (\ref{eqHint}), and note
that the polarization $P$ of the medium for the probe transitions can be expressed as a
partial derivative of the average single-atom interaction energy
with respect to the electric field
or, respectively, the Rabi-frequencies $E_i$
\begin{equation}
P_i=-\frac{N d_i}{\hbar}
\left\langle \frac{\partial {H}_{\rm int}}{\partial E_i^*}\right\rangle
\, {\rm e}^{-i\nu_i (t- z/c)}+c.c.,\quad i=1,2
\end{equation}
A similar expression holds for the drive field polarizations.
Here $\left\langle ...\right\rangle $ denotes quantum-mechanical
averaging, $d_i$ the dipole matrix elements of the corresponding transition 
and $N$ is the atom density. Hence
one can directly obtain the stationary field equations
in slowly-varying amplitude and phase approximation:
\begin{equation}
\frac{{\rm d} E_i}{{\rm d} z}=-i \frac{\eta_i}{\hbar}
\left\langle \frac{\partial {H}_{\rm int}}{\partial E_i^{\ast }}
\right\rangle ,  \label{Max2}
\end{equation}
where $\eta_i = N d_i^2 \omega_i/(2\hbar c \epsilon_0)$.
This approach allows us to obtain
the polarizations and equations of motion for the fields without
calculating the density matrix \cite{hamiltonian}. 

Using (\ref{Max2}), going to a comoving
frame via the transformation $(z,t)\to (\zeta=z,\tau=t-z/c)$,
assuming that all dipole moments (averaged over orientations
of the atoms) are approximately the same 
and introducing
the common coupling coefficient
$\kappa= N \overline{\, d_i^2\,} \omega_i/(2\hbar c\epsilon_0) =
3N\lambda^2\gamma /8 \pi$, gives the
following equations of motion for the fields:
\begin{eqnarray}
\frac{\partial}{\partial\zeta} E_1 &=& -i\kappa
\frac{\Omega_1^\ast\Omega_1^2 \Omega_2 E_2^\ast
- E_1^2  E_2  \Omega_1^\ast \Omega_2^\ast}
{\Delta \left(|\Omega_1|^2 + |E_1|^2\right)^2}, \label{eqE1} \\
%
\frac{\partial}{\partial\zeta} E_2 &=& -i\kappa
\frac{\Omega_1 \Omega_2 E_1^\ast}
{\Delta\left(|\Omega_1|^2+|E_1|^2\right) }, \\
%
\frac{\partial}{\partial\zeta} \Omega_1 &=&i\kappa
\frac{\Omega_1^2\Omega_2 E_1^*E_2^*-|E_1|^2E_1 E_2\Omega_2^*}
{\Delta \left(|\Omega_1|^2+|E_1|^2\right)^2}, \\
%
\frac{\partial}{\partial\zeta} \Omega_2 &=& -i\kappa
\frac{E_1 E_2\Omega_1^*}
{\Delta \left(|\Omega_1|^2+|E_1|^2\right)}. \label{eqW2}
\end{eqnarray}
Without sacrificing the underlying physics we can assume that the intial 
intensities of the pump fields and the intial intensities of the generated 
fields are equal:
\begin{equation}
|\Omega_1|^2 = |\Omega_2|^2, \;\;\;\; |E_1|^2 = |E_2|^2.
\end{equation}
As can easily be seen, the 
constants of motion imply that asymmetric initial conditions 
lead to exactly the same dynamics with the intensities merely shifted
up or down by a constant.

\begin{figure}[ht]
\begin{center}
\includegraphics[width=6.5cm]{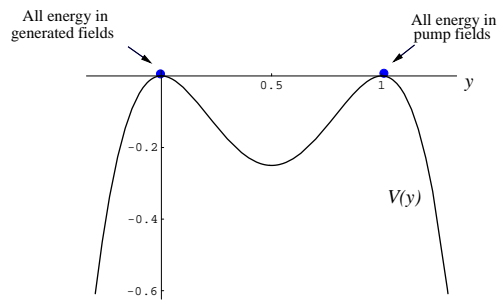}
\caption{Nonlinear pendulum potential experienced by a ``particle''
corresponding to the exchange energy between the pump and generated
fields. $\kappa/\Delta$ taken to be unity.} 
\label{figNonLinearPendulumPotential}
\end{center}
\end{figure}

Introducing a normalized intensity $y(t)$ with the identification 
$\zeta\to t$
\be
y(t) &=& \frac{|\Omega_1 (t)|^2}{|\Omega_1|^2+|E_1|^2},\\
 \dot y(t) &=&
-2 \frac{\kappa}{\Delta} \,\frac{{\rm Im}
\bigl(\Omega_1\Omega_2 E_1^* E_2^*\bigr)}{\bigl(|\Omega_1|^2+|E_1|^2\bigr)^2}
\ee
and making use of the constants of motion (\ref{eqCOM1}) -- (\ref{eqCOM3}) to 
reduce the problem to one variable, we obtain the differential equation
\begin{equation}
\frac{(\Delta/\kappa)^2}{2}\frac{{\rm d^2} y}{{\rm d }t^2} = 
\frac{\rm d}{{\rm d} y}
\Bigl(4 y^2 (y-1)^2\Bigr).
\end{equation}
In this form it is clear that the normalized intensity of the drive field 
can be identified with a particle of mass $m=\Delta/\kappa$ 
moving in the potential $V(y) =  -4y^2 
(y-1)^2$, depicted in Figure
\ref{figNonLinearPendulumPotential}. Conservation of energy requires  
that the particle be trapped between $0 \leq y \leq 1$. If $E_1(0) = E_2(0) 
= 0$, i.e. there is no seeding of the generated fields, then we have the 
case where $y(0)=1$. As is clear from the potential diagram, this 
corresponds to a point of unstable equilibrium, and indicates that no 
dynamics can take place. 

Choosing to seed the generated fields with a 
specific amplitude and phase corresponds to moving the particle off the 
critical point in the potential diagram. The drive field intensity will then 
oscillate, transferring energy back and forth between the drive and 
generated fields according to the constants of motion. The dynamics is
however sensitive to both the strength of the seed fields and the
initial relative phase between all four fields. The exact analytical solution
shows that the oscillation period of the energy transfer as well
as the maximum conversion efficiency depend on the seed-field intensity and
phase \cite{korsunsky1999,johnsson2002}.

%
%


\section{Quantum theory of stationary forward four-wave mixing}



\subsection{Exact numerical solution to the stationary quantum problem}


While forward four-wave mixing is relatively well understood
in the semiclassical case, very
little work has been done on the fully quantum case, that is, where
both the atomic subsystem and the interacting electromagnetic fields
are quantized. There are at least three reasons for considering such a
formulation.

First, from (\ref{eqE1}) -- (\ref{eqW2}) we see that the nonlinear
interaction of the system 
actually increases as the strength of the pump fields is
reduced. However, when the strength is reduced to a level
corresponding to only a few photons, the semiclassical approximations
must fail. Nonetheless, the extreme nonlinearities experienced in
conditions of very weak fields opens the intriguing possibility of
nonlinear effects at the few-photon level.

Secondly, a semiclassical analysis is not able to cope with the physically
realistic situation where initially only two fields are
present. Vacuum seeding through quantum fluctuations cannot be
described in the semiclassical framework, and one must always 
introduce at least a third field into
the analysis so that conversion can take place.

Thirdly, the behaviour of the system, at least 
in a semiclassical
analysis, is extremely sensitive to variations in intial phase and
amplitude of the pump and seed fields. When the fields are treated
quantum mechanically, these parameters will often be indeterminate,
particularly when starting from a vacuum field.

While an analytic solution of the fully quantum case appears
intractable, it is nevertheless possible to obtain exact solutions
numerically under stationary conditions. The stationary spatial 
evolution of the fields can be described by considering the time 
evolution of four harmonic oscillators $\hat a_1, \hat a_2$ and
$\hat b_1, \hat b_2$ corresponding to the generated and pump
fields respectively, interacting via the nonlinear Hamiltonian
\be
\hat H_{\rm int} = \frac{\hbar\kappa c}{\Delta} \biggl[
\frac{\hat b_1^\dagger \hat b_2^\dagger \hat a_1 \hat a_2
+ \hat a_1^\dagger \hat a_2^\dagger \hat b_1 \hat b_2}{
\hat b_1^\dagger\hat b_1+\hat a_1^\dagger \hat a_1}\biggr]
\label{eqHint-single}
\ee
with the following identification $\hat E_i\to \hat a_i$, $\hat\Omega_i\to \hat b_i$ and $\kappa=3N\lambda^2\gamma /8 \pi$.

The technique we use is to describe the possible states of
the system in a number state basis of the fields, use the 
effective Hamiltonian (\ref{eqHint-single}) 
and solve the resulting Schr\"{o}dinger equation
numerically. We will use the notation $| n_{\Omega_1} \; n_{\Omega_1}
\; n_{E_1} \; n_{E_2} \rangle$ for our basis vectors, 
where $n_{\Omega_1}$ is the number of
photons in the $\Omega_1$ mode, $n_{\Omega_2}$ is the number of
photons in the $\Omega_2$ mode and so on.  

The difficulty with this technique is that even though we have
eliminated the 
atomic degrees of freedom by using an effective Hamiltonian, because
we are considering a four-wave 
mixing process the size of the Hilbert space will scale as $n^4$, where
$n$ is a characteristic number of photons in each of the
fields. Consequently, if we wished to consider, say, 100 photons in
each of the beams, our Hilbert space would be $10^8$ dimensional, and
the problem would involve diagonalizing matrices with $10^{16}$
elements. Thus, as it stands, this approach is not computationally feasible.

The scaling problem can be avoided by using the constants of motion
(\ref{eqCOM1}) -- (\ref{eqCOM3}). Taken together, these relations allow
us to reduce a problem with four degrees of freedom to just one,
which is essentially the energy transfer from one field to
another. For example, (\ref{eqCOM2}) states that when a photon is
annihilated in the $\Omega_1$ mode, another must be created in the $E_1$
mode. 
Reduction of the problem to one with a single degree of freedom allows
us to choose the basis
\begin{equation}
\Psi_{nm} = \bigl| n_1 - n \;\,\, n_2-n \;\,\, n_3 + n \;\,\, n_4 + n\bigr\rangle
\end{equation}
where $n_1$, $n_2$, $n_3$ and $n_4$ denote the number of photons
initially in each of the four modes, and $n$ is the single degree of
freedom denoting how many 
photons have been transferred out of the pump mode. 


\subsection{Intensity evolution and 
quantum limitation of conversion efficiency}
\label{quantum-evolution}

\begin{figure}[ht]
\begin{center}
\includegraphics[width=7.5cm]{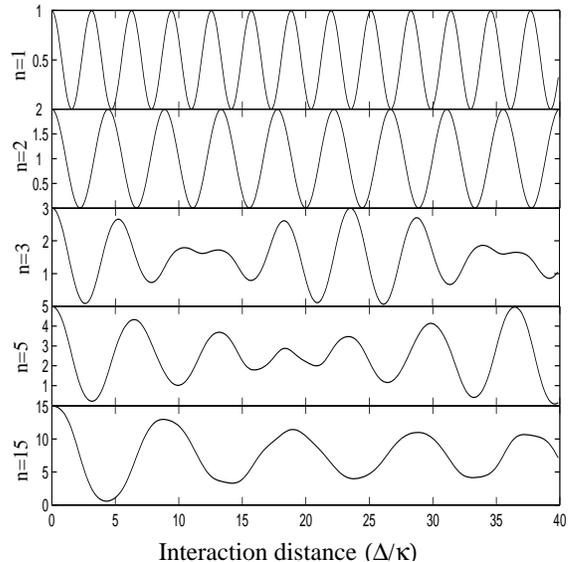}
\caption{Evolution of average pump photon number from
initial number
states. From top to bottom the initial photon number in the
pump modes are 1, 2, 3, 5 and 15 photons.} 
\label{figLowFockStates}
\end{center}
\end{figure}

We first consider the evolution of the fields when the initial states of the 
pump fields
are given by number states. This situation is computationally easy,
and we have calculated results for initial states
consisting of up to several thousand photons. A number state consisting of
several thousand photons, however, is not particularly experimentally
realistic, and thus here we consider only low photon numbers. The
results are shown in Figure \ref{figLowFockStates}. 

We note that the solutions are oscillatory, in qualitative 
analogy to the semiclassical predictions. 
In the fully quantum case, however, we see that only when
the initial pump fields contain one or two photons are the
oscillations comprised of just one frequency. If the initial fields
contain three or more photons several oscillation frequencies are
present, leading to a reduced conversion
efficiency. Provided long enough interaction distances are considered,
however, the different oscillation frequencies come back into
phase and reinforce each other, leading to the conclusion that
complete conversion can always be obtained at some point.

Another point to note is that as the intensity of the initial pump
fields is increased, the oscillation period increases. This is in
agreement with the form of the Hamiltonians (\ref{eqHint}), 
(\ref{eqHint-single}). The
denominator is a constant of motion and is related to the intensity of
the pump field $\Omega_1$. Thus a more intense pump field gives a
smaller interaction energy and consequently a longer conversion
distance.

\begin{figure}[ht]
\begin{center}
\includegraphics[width=7.5cm]{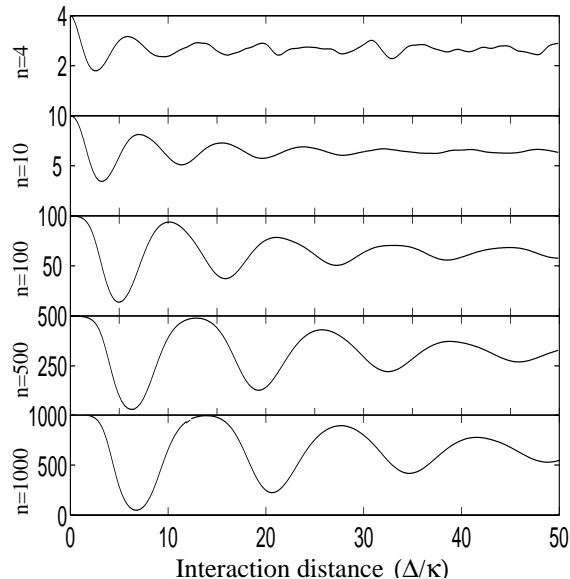}
\caption{Evolution of average pump photon number
for coherent initial states. From top
to bottom: an average photon number of 4, 10, 100, 500 and 1000.} 
\label{figCoherentStates}
\end{center}
\end{figure}

Next we consider the case where the pump fields are initially
described by coherent states. This situation is considerably more
computationally intensive, but calculations with an average photon number
of up to 1000 in each pump mode have been carried out. The results are
shown in Figure \ref{figCoherentStates}.
%
%
There are several overall qualitative features.

We see oscillations on a short distance
scale damping out over a longer distance scale to a conversion
efficiency of approximately one third. The damping at longer distances
can be explained by considering a coherent state to be superposition
of number states. Each number state has a different oscillation
period, and consequently interfere with each other and get out of
phase. As would be expected, on still longer distance scales
fractional revivals are seen.
As the input power is increased, the conversion distance (distance to the
first mininum in pump field intensity) increases
logarithmically while the conversion efficiency asymptotically
approaches unity. The scaling of the conversion distance with input power
in the resonantly enhanced four-wave mixing scheme is exactly opposite to 
the case of ordinary off-resonant four-wave mixing. There the 
conversion distance decreases with increasing input power.


\subsection{Quantum correlations and fluctuations of the generated fields}

\begin{figure}[ht]
\begin{center}
\includegraphics[width=6cm]{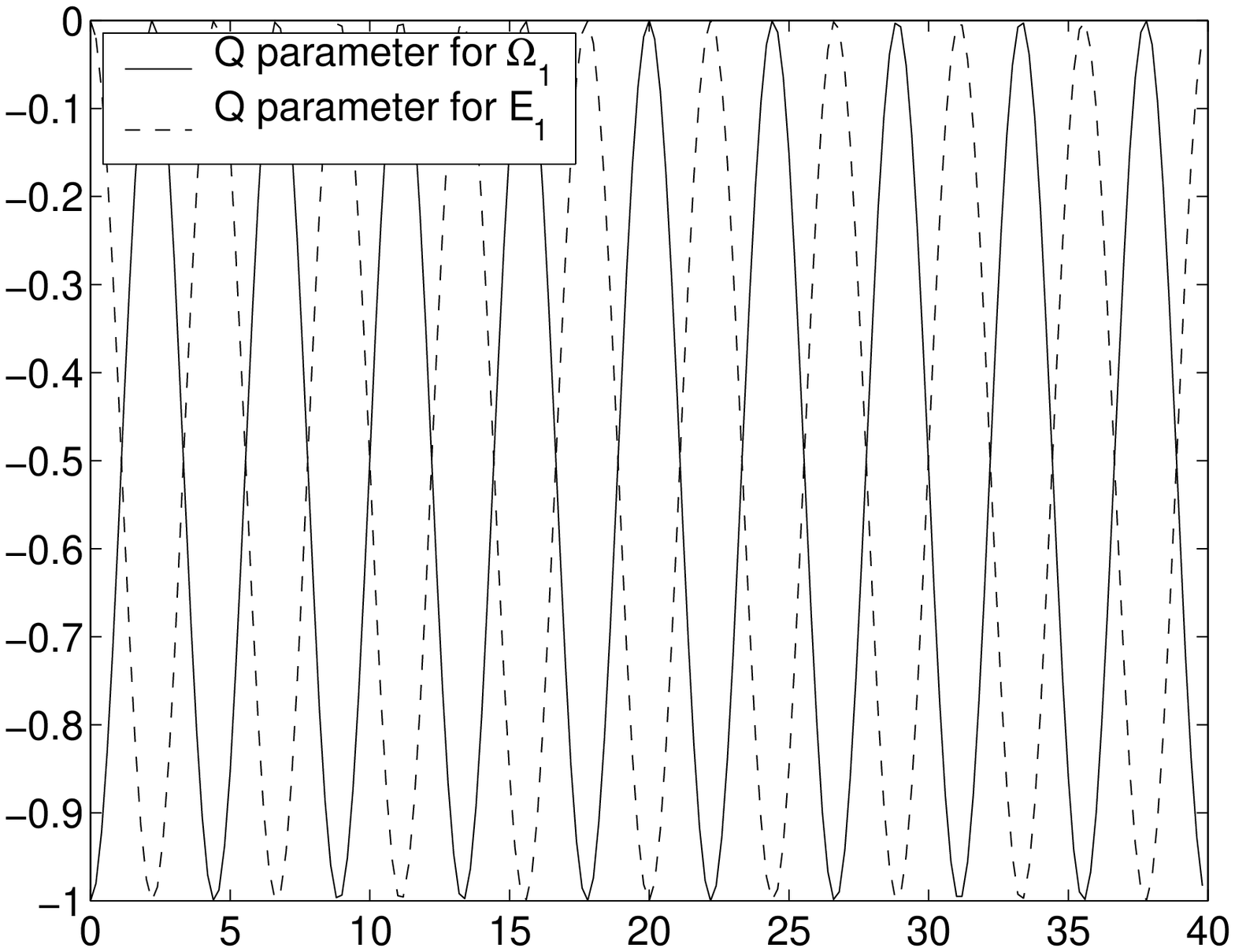}
\includegraphics[width=6cm]{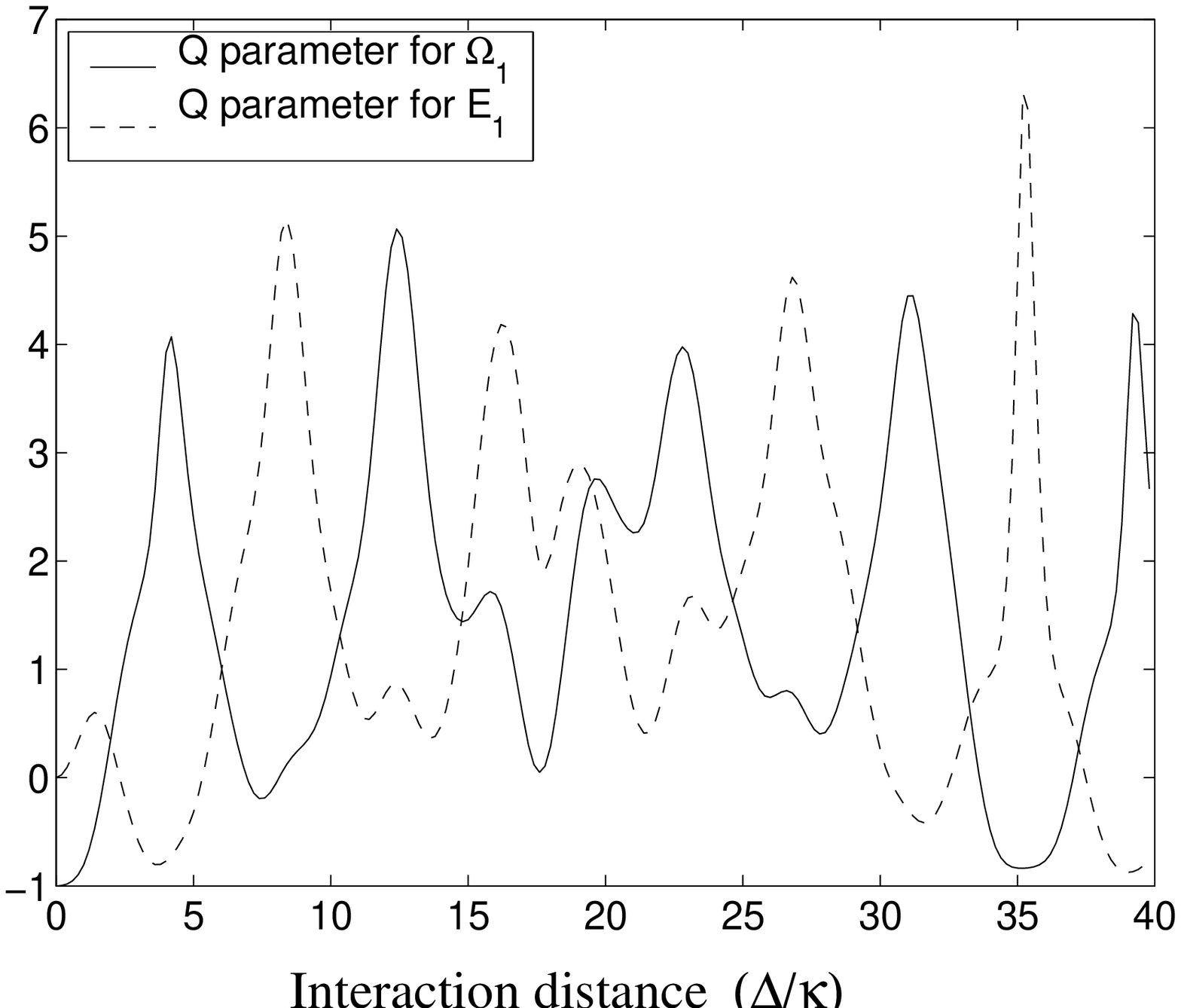}
\caption{Mandel $Q$ parameter for initial Fock states with $n=2$ (top)
and $n=10$ (bottom) in both
pump modes. Dashed line shows $Q$ parameter for generated fields, full line
for pump fields.} 
\label{Q-parameter-Fock-2}
\end{center}
\end{figure}

We now discuss the quantum fluctuations and correlations of the
generated fields. It is immediately evident from the interaction
Hamiltonian that when the two generated fields start
in the vacuum state, they will at all times be perfectly photon-number
correlated. Only states with equal photon number in both modes can
be generated. Consequently the intensity difference between the two fields
is perfectly squeezed:
\be
\left\langle \Delta\bigl(\hat a_1^\dagger\hat a_1
-\hat a_2^\dagger\hat a_2\bigr)^2
\right\rangle \equiv 0.
\ee
To characterize the statistics of photon pairs in the two generated modes
we have calculated the $Q$-parameter for different input states and 
intensities. As can be seen from Figure \ref{Q-parameter-Fock-2} 
 the pair statistics remain sub-Possonian
($Q\le 0$) for a Fock-state input with $n\le 2$. For a Fock-state
input with $n>2$ the pair statistics have a sub-Poissonian character only
for very small interaction distances 
and around the revival of the input intensity (see Fig.
\ref{figLowFockStates}). 

\begin{figure}[ht]
\begin{center}
\includegraphics[width=6.4cm]{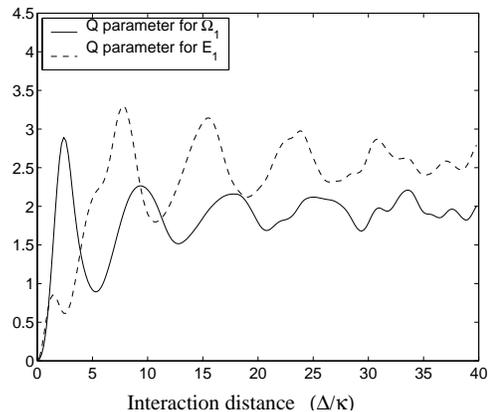}
\caption{Mandel $Q$ parameter for initial coherent states
 with $\langle n\rangle=10$ in both
pump modes. Dashed line shows $Q$ parameter for generated fields, full line
for pump fields.
} 
\label{Q-parameter-Coherent-10}
\end{center}
\end{figure}

For an initial coherent state the pair statistics remain
super-Poissonian at all times. This is illustrated in Figure 
\ref{Q-parameter-Coherent-10}.
%
%


\subsection{Realization of phase gate for continuous-variable
 quantum computation}


The field evolution from an initial Fock state 
with $n=1$ photons in each of the pump modes, discussed in subsection
\ref{quantum-evolution}, exhibits an interesting feature. After a
single complete cycle of energy conversion to the generated fields and
back, the quantum state of the system undergoes a phase change of
exactly $\pi$. This can be used to implement a phase gate for
continuous quantum computation.  Noting that no
conversion occurs unless both input modes are excited and fixing the
medium length at a value corresponding to twice the conversion length
for a single photon input of both pump modes one has the following
evolution of states 
\be
|0,0,0,0\rangle &\longrightarrow &|0,0,0,0\rangle ,\nn\\
|1,0,0,0\rangle &\longrightarrow &|1,0,0,0\rangle ,\nn\\
|0,1,0,0\rangle &\longrightarrow &|0,1,0,0\rangle ,\\
|1,1,0,0\rangle &\longrightarrow &-|1,1,0,0\rangle .\nn
\ee
This is a perfect realization of a phase gate.


\section{Mean-field theory of resonant forward four-wave mixing}


In order to obtain a better understanding of the quantum dynamics of resonant
four-wave mixing, in particular the limitations and scaling of conversion
efficiency as well as the conversion distance, 
we now  derive approximate analytical solutions
within an appropriate mean-field theory. 
Our starting point is again the effective Hamiltonian (\ref{eqHint-single}), 
along with the assumption that we can replace the denominator, a constant of 
motion, with its expectation value. In this case the equations of
motion for the fields in the Heisenberg picture are
\begin{eqnarray}
\frac{\rm d}{{\rm d} t}\hat b_1 &=& -\frac{i\kappa c}{\Delta d} \;
\hat b_2^{\dagger} \hat a_1 \hat a_2  \label{eqMFEOM1} \\
\frac{\rm d}{{\rm d} t}\hat a_1 &=& -\frac{i\kappa c}{\Delta d} \;
\hat a_2^{\dagger} \hat{b}_1 \hat{b}_2  \label{eqMFEOM2} \\
\frac{\rm d}{{\rm d} t} \hat b_2
&=& -\frac{i\kappa c}{\Delta d} \;
\hat b_1^{\dagger} \hat{a}_1 \hat{a}_2  \label{eqMFEOM3} \\
\frac{\rm d}{{\rm d} t} \hat a_2
&=& -\frac{i\kappa c}{\Delta d} \;
\hat{a}_1^{\dagger} \hat{b}_1 \hat{b}_2 \label{eqMFEOM4}
\end{eqnarray}
where $d=\langle\hat b_1^{\dagger}\hat b_1 + 
\hat{a}_1^{\dagger}\hat{a}_1\rangle$.

The equations of motion for the average field intensities
$\langle \hat b_1^\dagger \hat b_1\rangle$ and so on will contain
four-field correlation functions, for example
\begin{equation}
\frac{\rm d}{{\rm d} \zeta} \langle \hat b_1^{\dagger} \hat b_1 \rangle = 
\frac{i\kappa}{\Delta d}\left[ \langle \hat b_1 \hat b_2 \hat a_1^{\dagger} 
\hat a_2^{\dagger} \rangle -  \langle \hat a_1 \hat a_2 \hat b_1^{\dagger} 
\hat b_2^{\dagger} \rangle \right]
\end{equation}
where we have once again switched to a spatial rather than a temporal
picture. 

To proceed further we  assume that all fields are gaussian and
that correlations between pump and generated fields can be neglected.
This assumption is reasonable as long as the pump fields are initially
in coherent states with sufficiently large amplitude.
With this decorrelation approximation, 
the fourth order expectation values can be expressed as sets
of bilinear terms of the form
$\langle\hat b_i^{\dagger} \hat b_j\rangle$ and 
$\langle\hat a_i^{\dagger} \hat a_j\rangle$. It is important that we
also
keep anomalous correlations, such as $\langle\hat a_i \hat a_j\rangle$,
as both pairs of fields are strongly correlated.
In addition, we can use the constants of motion (\ref{eqCOM1}) --
(\ref{eqCOM3}) to relate the expectation values of the number operators
$\langle \hat b_1^{\dagger}\hat b_1\rangle$, $\langle
\hat b_2^{\dagger}\hat b_2\rangle$, $\langle \hat
a_1^{\dagger} \hat a_1\rangle$ and $\langle \hat a_2^{\dagger} 
\hat a_2\rangle$.

If we define $\langle\hat b_1 \hat b_2\rangle =
b_{12} e^{i\varphi_{b}}$,  $\langle\hat a_1 \hat a_2\rangle =
a_{12} e^{i\varphi_{a}}$, and  $\langle\hat b_1^{\dagger} 
\hat b_1\rangle =b$, 
then we obtain the following set of coupled, complex-valued
nonlinear differential equations:
\begin{eqnarray}
\frac{\rm d}{{\rm d}\zeta}
a_{12} &=& \frac{\kappa}{\Delta d}(2a +1) a_{12} \sin(\varphi_a -
\varphi_{b}) \label{eqOmega12} \\
 \frac{\rm d}{{\rm d}\zeta}a_{12} &=& 
\frac{\kappa}{\Delta d}(2 b -2d -1) b_{12} \sin(\varphi_a -
\varphi_{b}) \\
\frac{\rm d}{{\rm d}\zeta} b &=& 
\frac{2\kappa}{\Delta d}b_{12} E_{12} \sin(\varphi_a -
\varphi_{b}) \\
\frac{\rm d}{{\rm d}\zeta}\varphi_{b} &=& 
-\frac{\kappa}{\Delta d}(2b +1)
\frac{a_{12}}{b_{12}} \cos(\varphi_a -
\varphi_{b}) \\
 \frac{\rm d}{{\rm d}\zeta}\varphi_{a} &=& 
\frac{\kappa}{\Delta d}(2b -2d-1)
\frac{b_{12}}{a_{12}} \cos(\varphi_a -
\varphi_{b}) \label{eqphiE}.
\end{eqnarray}
To simplify this system we note that if we take expectation values
 of both sides of
(\ref{eqCOM4}) we find $b_{12} a_{12} \cos(\varphi_a -
\varphi_{b})$ is a constant of motion. If we assume that both
generated fields start from vacuum this constant is equal to
zero, and since $b_{12}$ and $a_{12}$ are not always zero we see
that $\cos(\varphi_{a}-\varphi_{b}) = \pm 1$, with the sign flip
coming at the end of each conversion cycle. This observation, in
conjunction with two more constants of motion that can be extracted from
(\ref{eqOmega12}) -- (\ref{eqphiE}), enable us to reduce the set of
five coupled equations to just one:
\begin{eqnarray}
\frac{{\rm d} b}{\rm d \zeta} =\frac{2\kappa}{\Delta b_0}
\sqrt{(b^2 + b - b_0)
(b-b_0)(b-b_0-1)}, \label{eqMeanFieldOmega} 
\end{eqnarray}
where $b_0 = \langle\hat b_1^{\dagger}(0)\hat b_1(0)\rangle =
\langle\hat b_2^{\dagger}(0) \hat b_2(0)\rangle$. This is the equation
governing the evolution of the expectation value of the number of
photons in each 
of the pump modes, under the assumption that both pump fields have the
same intensity and the two generated fields start from vacuum.

\begin{figure}[ht]
\begin{center}
\includegraphics[width=7cm]{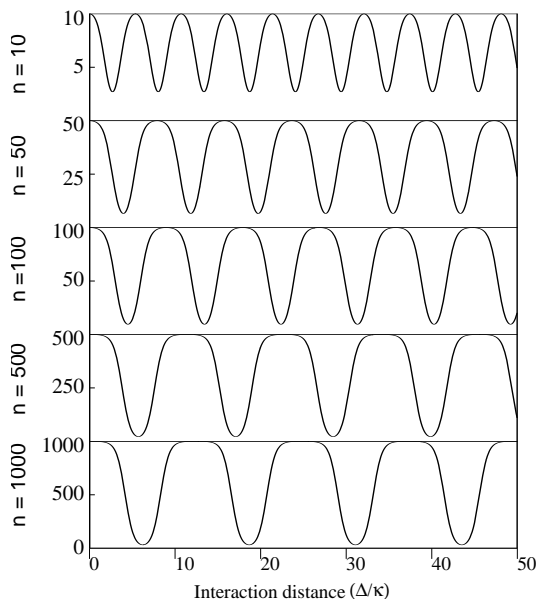}
\caption{Pump field intensity in the mean-field approximation. From
top to bottom: 10, 50, 100, 500 and 1000 photons.
$\kappa/\Delta=1$, $b_0=10$.} 
\label{figMeanFieldSolutions}
\end{center}
\end{figure}

\begin{figure}[ht]
\begin{center}
\includegraphics[width=7cm]{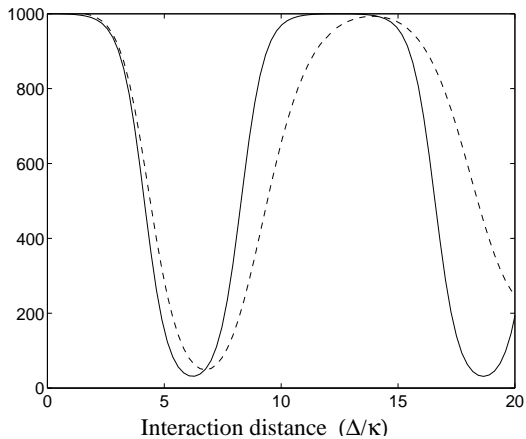}
\caption{Comparison between the exact numerical solutions (dashed
line) and the mean-field theory (solid line) for a coherent state
with an average photon number of 1000.} 
\label{figMeanfieldVsNumericalComparision}
\end{center}
\end{figure}

As a fourth order polynomial is involved, this
differential equation can be (implicitly) solved analytically in 
terms of elliptic
integrals, but the specific form of the solution is involved and 
not particularly
illuminating. Numerical solutions to (\ref{eqMeanFieldOmega}) are
plotted in Figure \ref{figMeanFieldSolutions} for a number 
of initial pump field
intensitities. 
%
%
One recognizes an oscillatory exchange of energy with 
a non-perfect conversion effiency. Comparison of the mean-field
results with the fully quantum calculation for the case of a coherent
pump input shows good agreement over the first period of energy exchange 
between pump and generated fields
(Figure~\ref{figMeanfieldVsNumericalComparision}). In particular the 
maximum conversion efficiency and the conversion distance are
well reproduced. As in the quantum solutions, the conversion period
increases logarithmically with the input intensity.
For larger distances the mean-field solution remains
periodic, while the oscillations in the true quantum case decay.
As the interaction distance increases,
higher order correlations build up and thus the gaussian factorization
approximation used in the mean-field theory breaks down.
%
%

Analoguous to the semiclassical case, (\ref{eqMeanFieldOmega}) can be
mapped to a nonlinear pendulum problem, with $b$
corresponding to the position of a particle with mass $\Delta^2/\kappa^2$
moving in a potential
\begin{equation}
V(b) = -\frac{2}
{b_0^2}(b^2 + b - b_0)(b-b_0)(b-b_0-1)
\label{eqMeanFieldPendulum}
\end{equation}
where $b$ plays the role of the particle position and the identification
$\zeta\to t$ is used.

\begin{figure}[ht]
\begin{center}
\includegraphics[width=6cm]{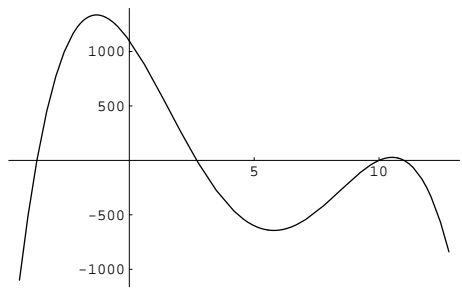}
\caption{Nonlinear pendulum potential experienced by a ``particle''
whose position corresponds to the number of photons in the pump modes.
$\kappa/\Delta=1$, $b_0=10$.} 
\label{figMeanFieldPendulumPotential}
\end{center}
\end{figure}

The potential (\ref{eqMeanFieldPendulum}) is plotted in Figure
\ref{figMeanFieldPendulumPotential} for the case $b_0 = 10$,
$\kappa/\Delta =1$.
%
%
From the shape of the potential one can see that different dynamics is
expected compared with the semiclassical case, where the potential is
described by Figure \ref{figNonLinearPendulumPotential}. In the
semiclassical analysis, choosing the initial intensity of 
$a_1$ and $a_2$ to be zero corresponds to
starting exactly on the right hand peak of the potential in Figure
\ref{figNonLinearPendulumPotential}, and consequently no dynamical
evolution can occur. 

\begin{figure}[ht]
\begin{center}
\includegraphics[height=5cm,width=6.5cm]{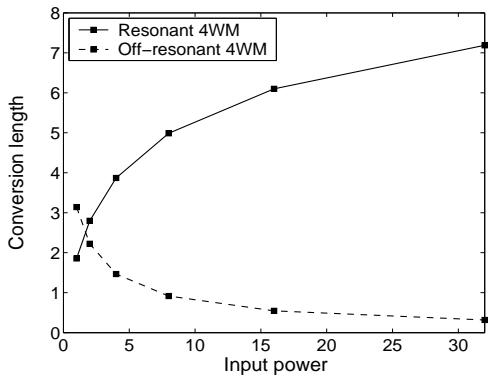}
\caption{Scaling of conversion distance as function 
of input intensity of coherent
pump for ordinary off-resonance four-wave mixing (dashed line, $d=$ const.)
and resonant four-wave mixing (full line).} 
\label{figComparisionConversionLength}
\end{center}
\end{figure}

In the mean-field theory, however, the starting position ($b_0)$
is always
slightly to the left of the right summit (peak
position is $\sim b_0 + \frac{1}{2}$ for large $b_0$), and
consequently oscillation will always take place. As the input power is
increased the starting position asymptotically approaches the right
hand peak, but never reaches it. 
From the potential (\ref{eqMeanFieldPendulum}) one can easily obtain the
conversion distance or oscillation period:
\be
z_{\rm conv}=\int_{b_0}^{b_1}\!\!\frac{{\rm d}b}{
\sqrt{-\frac{2\Delta}
{\kappa} V(b)}}
\ee
where $b_1$ is the inner turning point. 
As the input power is increased, it
takes longer for the oscillation to begin due to the flatter gradient
of the potential, leading to the logarithmic increase in conversion period.
This is in sharp contrast to ordinary off-resonance four-wave mixing, where
the conversion length decreases with input power. The effective hamiltonian
discussed in the present paper can be converted into that of ordinary
off-resonant four-wave mixing by replacing the intensity dependent
denominator $d$ by a constant. Figure \ref{figComparisionConversionLength}
gives a comparision of the conversion distance as a function
of input power for the two cases. The periods were found from
numerical solutions to the full quantum problem using the Hamiltonian
(\ref{eqHint}) with and without the resonant denominator.
%
%
 One clearly recognizes the peculiar
feature of the resonant process to work better for smaller input intensities.

From the roots of (\ref{eqMeanFieldPendulum}) 
one can immediately obtain a result connecting 
input intensity and conversion efficiency. Defining the conversion
efficiency $e = b_{\textrm{min}}/b_0$ we find
\begin{equation}
e=1-\frac{1}{2b_0}(\sqrt{4b_0+1} -1)
\sim 1- \frac{1}{\sqrt{b_0}}.
\end{equation}
Thus in the mean-field model the conversion efficiency $(1-e)$ scales with
the input pump field  as $b_0^{-1/2}$.


\section{summary}


In the present paper we have discussed the quantum theory of stationary 
resonant four-wave mixing in a forward scattering configuration. We
considered the interaction of four electromagentic fields with atoms
in a modified double-$\Lambda$ configuration. The modified five-state
coupling scheme was used instead of the standard four-level one
in order to eliminate ac-Stark induced nonlinear phase
shifts \cite{johnsson2002}. This is necessary, because a full quantum
analysis shows that full conversion cannot be obtained when these
terms are present, even in the asymptotic limit --- 50\% conversion is
the best that can be achieved using coherent states. 

To eliminate the atomic variables we assumed adiabatic following and
derived a classical nonlinear interaction 
Hamiltonian of the four electromagnetic fields. In contrast to ordinary
off-resonant four-wave mixing, the denominator of this Hamiltonian
contains 
the sum of the intensities of the resonant pair of pump and generated
fields. This is a consequence of the infinitely long-lived two-photon 
resonance which is entirely determined by power-broadening for
arbitrarily small field intensities. As a consequence the effective
Hamiltonian cannot be expanded as
a power series in the fields and the nonlinear interaction becomes stronger
the smaller the pump field intensity. The classical evolution can be mapped
onto a one-dimensional nonlinear pendulum and solved exactly, provided
small seed fields are included in the analysis. The initial
evolution of the generated fields from vacuum, however, is entirely
determined by quantum fluctuations and thus cannot be described within
a classical approach.  

To obtain a full quantum solution under stationary conditions
we quantized the effective interaction Hamiltonian by replacing c-number
expressions by properly ordered operator products. We showed that the
constants of motion of the Hamiltonian allow us to reduce the stationary
wave mixing process to a single-mode problem, which can be solved numerically
for up to $10^3$ input photons. We have analyzed the
field evolution starting from initial Fock and coherent states in the pump
modes. In both cases an oscillatory energy exchange between the modes
was found. For the case of initial Fock states with up to $n=2$ photons
in both pump modes, the oscillation is purely sinusoidal
with a single frequency and complete conversion is possible. For higher
input photon numbers multiple frequencies appear and complete conversion
is only possible after several oscillation periods. In the case of coherent
input fields the oscillations in the energy exchange are damped and complete
conversion can be achieved only asymptotically for large input intensities. 

As the four-wave mixing process simultaneously creates photons in two modes, 
the photon numbers in those modes are perfectly quantum correlated,
which has potential applications in quantum communication systems.
The statistics of the photon pair production is, however, mostly 
super-Poissonian. 

We observed an interesting property of the four-wave mixing process when
the two input fields are both in a single-photon Fock state. After a complete
conversion cycle into the generated fields and back the quantum state
attains a phase shift of exactly $\pi$. This allows one to construct an
ideal phase gate for continuous variable quantum computation. 

In order to gain more physical insight in the
quantum four-wave mixing we developed a mean-field theory assuming 
gaussian and independent pump and generated fields but taking into
account anomalous field correlations. The mean-field equations can again
be mapped to a one-dimensional anharmonic pendulum and solved exactly.
The solutions for the average intensities reproduce the
quantum results for coherent input over the first oscillation period.
In particular analytic expressions for the conversion length and
the conversion efficiency can be derived which coincide very well
with the exact results. 


\section*{Acknowledgement}


M.J. acknowledges financial support by the European Union research network COCOMO.

\vfill

\def\etal{\textit{et al.}}


\begin{thebibliography}{99}


\bibitem{harris1997} S. E. Harris, Physics Today {\bf 50}, 36 (1997).

\bibitem{lukin2001} M. D. Lukin and A. Imamoglu, Nature {\bf 413}, 273 (2001).

\bibitem{marangos1998} J. P. Marangos, 
Journal of Modern Optics {\bf 45}, 471 (1998).

\bibitem{Harris1990} S. E. Harris, J. E. Field, and A. Imamoglu, 
Phys. Rev. Lett. {\bf 64}, 1107 (1990).

\bibitem{Stoicheff1990}   K. Hakuta, L. Marmet, and B.P. Stoicheff, 
Phys. Rev. Lett.
\textbf{66}, 596 (1991); G.Z. Zhang, K. Hakuta, and B.P. Stoicheff, 
Phys. Rev. Lett.\textbf{71}, 3009 (1993).

\bibitem{jain1996}  M. Jain, H. Xia, G.Y. Yin, A.J. Merriam, and S.E. Harris,
Phys. Rev. Lett. \textbf{77}, 4326-4329 (1996); S.E. Harris, G.Y. Yin, M.
Jain, H. Xia, and A.J. Merriam, Phil. Trans. R. Soc. Lond. A \textbf{355},
2291-2304 (1997).






\bibitem{harris1998} S. E. Harris and Y. Yamamoto, Phys. Rev. Lett. {\bf 81},
3611 (1998).


\bibitem{harris1999} S. E. Harris and L. V. Hau, Phys. Rev. Lett. {\bf 82},
 4611 (1999).


\bibitem{imamoglu2000} M. D. Lukin and A. Imamoglu, Phys. Rev. Lett. 
{\bf 84},1419 (2000).




\bibitem{schmidt1996} H. Schmidt and A. Imamoglu, Opt. Lett. 
{\bf 21}, 1936 (1996).

\bibitem{hemmer1994} P.R. Hemmer, K. Z. Cheng, J. Kierstead, M. S. Shariar, 
and
M. K. Kim, Opt. Lett. {\bf 19}, 296 (1994); B. S. Ham, M. S. Shariar, M. K. 
Kim,
and P. R. Hemmer, Opt. Lett. {\bf 22}, 1849 (1997).


\bibitem{babin1996} S. Babin, U. Hinze, E. Tiemann, B. Wellegehausen, 
Opt. Lett. {\bf 21}, 
1186 (1996); S. Babin, E. V. Podivilov, D.A. Shapiro, U. Hinze, 
E. Tiemann, B. Wellegehausen,
Phys. Rev. A {\bf 59}, 1355 (1999); U. Hinze, L. Meyer, B.N. Chichkov, 
E. Tiemann,
B. Wellegehausen, Opt. Comm {\bf 166}, 127 (1999).

\bibitem{lukin1997}
 M. D. Lukin, M. Fleischhauer, A.S. Zibrov, H.G. Robinson, V.L. Velichansky,
L. Hollberg, M.O. Scully, Phys. Rev. Lett. {\bf 79}, 2959 (1997).

\bibitem{popov1997} 
A. K. Popov, S. A. Myslivets, Kvant. Elektr. {\bf 24}, 1033 (1997).

\bibitem{lu1998}
B. Lu, W.H. Burkett, M. Xiao, Opt. Lett. {\bf 23}, 804 (1998).
 

\bibitem{korsunsky1999}
E.~A.~Korsunsky and D. V. Kosachiov,
Phys. Rev. A {\bf 60}, 4996 (1999).

\bibitem{korsunsky2002} E. A. Korsunsky and M. Fleischhauer, 
preprint quant-ph/0204089.

\bibitem{lukin1998} M.~D.~Lukin, P.~Hemmer, M.~Loeffler, and M.~O.~Scully, 
Phys. Rev. Lett. {\bf 81}, 2675 (1998).

\bibitem{review} M. D. Lukin, P. R. Hemmer, M. O. Scully, 
in {\it Adv. At. Mol. 
and Opt. Physics}, {\bf 42B}, 347 (Academic Press, Boston, 1999).  

\bibitem{fleischhauer2000book} M. Fleischhauer in {\it Frontiers of Laser 
Physics and Quantum Optics}, 
(Z. Xu, S. Xie, S.-Y. Zhu and M. O. Scully, Eds.), p. 97-106 (Springer, 
Berlin, 2000).

\bibitem{zibrov1999} A. S. Zibrov, M. D. Lukin and M. O. Scully, 
Phys. Rev. Lett. {\bf 83}, 4049 (1999). 

\bibitem{yuen1979} H. P. Yuen and J. H. Shapiro, Opt. Lett. {\bf 4},
334 (1979).

\bibitem{bloembergen1965} N. Bloembergen, {\it Nonlinear Optics} 
(W. A. Benjamin, New York 1965).

\bibitem{boyd1992} R. W. Boyd,  {\it Nonlinear Optics} 
(Academic Press, San Diego 1992).

\bibitem{lukin1999} M. D. Lukin, A. B. Matsko, M. Fleischhauer, 
and M. O. Scully, Phys. Rev. Lett. {\bf 82}, 1847 (1999) 

\bibitem{fleischhauer2000} 
M. Fleischhauer, M. D. Lukin, A. B. Matsko, and M. O. Scully, 
    Phys. Rev. Lett. {\bf 84}, 3558 (2000).


\bibitem{johnsson2002} Mattias Johnsson, Evgeny Korsunsky and Michael 
Fleischhauer, preprint quant-ph/0205194

\bibitem{hamiltonian} the approach is adapted from:
  A. O. Melikyan and S. G. Saakyan, Zh. Exp.Teor. Fiz. \textbf{%
76}, 1530 (1979) [Sov. Phys. JETP \textbf{49}, 776 (1979)];
A. R. Karapetyan and B. V. Kryzhanovskii, Zh. Exp. Teor. Fiz.
\textbf{99}, 1103 (1991) [Sov. Phys. JETP \textbf{72}, 613 (1991)]; B.
Kryzhanovsky and B. Glushko, Phys. Rev. A \textbf{45}, 4979 (1992).



\end{thebibliography}
\end{document}